\documentclass[12pt]{article}
\usepackage{eucal,  amssymb,amsmath,graphicx, amsfonts, latexsym}
\usepackage[T1]{fontenc}
\usepackage[]{natbib}
\usepackage{verbatim}

\usepackage{authblk}
\providecommand{\keywords}[1]{\textbf{Keywords: } #1}

\usepackage{etoolbox}

\makeatletter
\patchcmd{\maketitle}{\@fnsymbol}{\@alph}{}{}  
\makeatother

\usepackage{url}

\begin{document} \parskip=5pt plus1pt minus1pt \parindent=0pt
\title{Introduction to statistical inference for infectious diseases}
\author[]{Tom Britton\thanks{ E-mail: tomb@math.su.se}  \ and Federica Giardina\thanks{ E-mail: federica@math.su.se}}
\affil[]{Department of Mathematics, Stockholm University, Stockholm, Sweden}
\date{\today}
\maketitle

\begin{abstract} 

In this paper we first introduce the general stochastic epidemic model for the spread of infectious diseases. Then we give methods for inferring model parameters such as the basic reproduction number $R_0$ and vaccination coverage $v_c$ assuming different types of data from an outbreak such as final outbreak details and temporal data or observations from an ongoing outbreak. Both individual heterogeneities and heterogeneous mixing are discussed. We also provide an overview of statistical methods to perform parameter estimation for stochastic epidemic models. In the last section we describe the problem of early outbreak detection in  infectious disease surveillance and statistical models used for this purpose. 

\end{abstract}


\section{Introduction}

Infectious disease models aim at understanding the underlying mechanisms that influence the spread of diseases and predicting disease transmission.  Modelling has been increasingly used to evaluate the potential impact of different control measures and to guide public health policy decisions. 

Deterministic models for infectious diseases in humans and animals have a vast literature, e.g.\ \cite{anderson1991infectious, keeling2008modeling}. Although these models can sometimes be sufficient to model the mean behaviour of the underlying stochastic system and guide towards parameter estimates, they do not allow the quantification of the uncertainty associated to model parameters estimates \citep{becker1989analysis}. Stochastic models \citep{Andersson2000,eltit,diekmann2012mathematical}, can be used to infer relevant epidemic parameters and provide estimates of their variability. 

Infectious disease data are commonly collected by surveillance systems at certain space and time resolutions. The main objectives of surveillance systems are early outbreak detection and the study of spatio-temporal patterns. Early outbreak detection commonly relies on statistical algorithms and regression models for (multivariate) time series of counts accounting for both time and space variations.

In this overview paper, we start by analysing the general stochastic epidemic model, which describe the spread of a Susceptible Infected Recovered (SIR) disease assuming a closed population with homogeneous mixing and describe how to make inference on important epidemiological parameters, namely the basic reproduction number  $R_0$ and the critical vaccination coverage $v_c$. We then describe inference procedures for various extensions increasing model realism. Moreover, we describe statistical models used for the analysis and forecasting of time series of infectious disease data in surveillance settings.

Section \ref{sec2} defines the general stochastic model, and describes inference procedures for $R_0$ and $v_c$ depending on the available data (final size or temporal data). Section \ref{sec3} presents extensions of the general stochastic models treating both individual and mixing heterogeneities and Section \ref{sec4} discusses the main issues in statistical inference from ongoing outbreaks, relating estimates of the exponential growth rate $r$ to $R_0$ using e.g.\ serial intervals and generation time estimation. The main challenge in parameter estimation for epidemic models is that the infection process is usually not observed. Section \ref{sec5} presents an overview of statistical methods to estimate transmission model parameters dealing with the missing data and describes recent advances in statistical algorithms to improve computational performance. Section \ref{sec6} shows how statistical models with space/time structures can be applied to infectious disease surveillance settings for early outbreak detection and forecasting. Section \ref{sec7} mentions some further extensions and model generalizations as well as new approaches to perform statistical inference for infectious diseases.

\section{Inference for a simple stochastic epidemic model}\label{sec2}

\subsection{A simple stochastic epidemic model and its data}

We start by defining a simple stochastic model known as the general stochastic epidemic model (e.g.\ Section 2.3 in \citet{Andersson2000}). This model considers a so-called SIR-disease where individuals at first are Susceptible. If they get infected they immediately become Infectious (an infectious individual is called an infective) and remain so until they Recover assuming immunity during the rest of the outbreak. Individuals can hence get infected at most once. The general stochastic epidemic assumes a closed population in which individuals mix uniformly in the community, and all individuals are equally susceptible to the disease and equally infectious if they get infected.

Consider a closed population of size $n$. An individual who gets infected immediately becomes infectious and remains so for an exponentially distributed time with rate parameter $\gamma$. During the infectious period an individual has ``close contact'' with other individuals randomly in time at rate $\lambda$, each such contact is with a uniformly selected individual, and a close contact is a contact which results in infection if the contacted person is susceptible; otherwise the contact has no effect.

Let $(S(t), I(t), R(t))$ denote the numbers of susceptible, infectious and recovered individuals at time $t$. Because the population is closed and of size $n$ we have $S(t)+I(t)+R(t)=n$ for all $t$. At the start of the epidemic we assume that $(S(0), I(0), R(0))=(n-1, 1,0)$, i.e.\ that there is one initially infective and no immune individuals. The model is Markovian implying that it may equivalently be defined by its jump rates. An infection occurs at $t$ with rate $\lambda I(t)S(t)/n$ (since each infective has contacts at rate $\lambda$ and a contact results in infection with probability $S(t)/n$). The other event, recovery, occurs at $t$ with rate $\gamma I(t)$, since each infective recovers at rate $\gamma$.

The epidemic evolves until the first (random) time $T$ when there are no infectives. Then both rates are $0$ and the epidemic hence stops. The final size of the epidemic is denoted $Z=R(T)$, the number of individual that were infected during the outbreak, all others still being susceptible ($S(T)=n-Z$).

The epidemic model has two parameters, $\lambda$ and $\gamma$, plus the population size $n$. The perhaps most important quantity for any epidemic model is called the \emph{basic reproduction number} and denoted $R_0$. The definition of $R_0$  is that it equals the average number of infections caused by a \emph{typical} individual during the early stage of an outbreak (when nearly all individuals are still susceptible). It is often defined assuming that the population size $n$ tends to infinity. For the general stochastic epidemic, the basic reproduction equals 
$$
R_0=\lambda/\gamma .
$$ 
This is so because an individual infects others at rate $\lambda$ (when all individuals are susceptible) while infectious, and the mean duration of the infectious period equals $1/\gamma$. The most important property of $R_0$ is that it has a threshold value at 1: if $R_0>1$, i.e.\ if infected individuals infect more than one individual on average, then the epidemic can take off thus producing a ``major outbreak'', whereas if $R_0\le 1$ the disease will surely die out without affecting a large fraction of individuals. This has important consequences for vaccination. If, prior to the outbreak, a fraction $v$ are vaccinated (or immunized in some other way), then the number of infections caused by a typical individual is reduced to $R_0(1-v)$ since only the fraction $1-v$ of all contacts result in infection. The new reproduction number is hence $R_v=(1-v)R_0$. For the same reason as above, a positive fraction of the community may get infected if and only if $R_v>1$. Using the expression for $R_v$ this is seen to be equivalent to $v>1-1/R_0$. The value $v_c$ where we have equality is denoted the \emph{critical vaccination coverage} and given by
$$
v_c=1-\frac{1}{R_0}.
$$
The conclusion is hence that the fraction necessary to vaccinate (or isolate in some other way)  to surely avoid a big epidemic outbreak is a simple function of $R_0$. This explains why $R_0$ and $v_c$ are considered the perhaps two most important parameters in infectious disease epidemiology (cf.\ \cite{anderson1991infectious}).

Now we study inference procedures for these parameters (and others) in the general stochastic  model. What we can infer, and with what precision, depends on the available data. Below we mainly focus on the two extreme types of data. The first is where we only observe the final size $Z=R(T)$. The second situation is where we have detailed information about the state of all individuals throughout the outbreak, i.e.\ where we observe the complete process $\{ (S(t), I(t), R(t));0\le t\le T\}$, called complete observation. In reality, it is often the case that some temporal information is available even if the exact state of all individuals is not known. For example, the onset of symptoms may sometimes be observed for infected individuals. How the onset of symptoms relate to the time of infection and time of recovery depends on the disease in question. Since we are not considering any specific disease, we treat the two extreme situations of final size and complete observation, the precision of any estimator based on partial temporal observations will lie between these two situations.

There are many extensions of the model defined above. For example, it is sometimes assumed that the infectious period is different from the exponential distribution assumed above. The situation where it is assumed non-random is called the continuous time Reed-Frost epidemic model, but also other distributions may be relevant. Another extension is where the disease has a latent period, i.e.\ where there is a period between when an individual gets infected and until he or she becomes infectious. Such models are often referred to as SEIR epidemics, where the ``E'' stands for ``exposed (but not yet infectious)''. Some perhaps even more important extensions are where the community is considered heterogeneous with respect to disease spreading. For example, some individuals (like children and elderly) may be more susceptible to the disease but it is also possible that certain individuals are more infectious be shedding more virus during the infectious period. A different form of heterogeneity of high relevance is where the community has heterogeneous social structures, which all communities do. For example, individuals are more likely to spread the disease to members of the same household than to a random individual in the community. 

There are two main reasons why making inference in infectious disease outbreaks is harder than in many other situations. The first is that infection events are not independent: whether I get infected is not at all independent of whether my friends get infected. Most standard theory for statistical inference is based on independent events, but such methods are hence not applicable in our situation. The second complicating factor is that we rarely observe the most important events: when and by whom an individual is infected and when they stop being infectious. Instead we observe surrogate observations such as onset of symptoms and stop of symptoms or similar, and to infer the former from the latter is not straightforward. Statistical methodology to analyse such data imputing missing observations will be reviewed in Section \ref{sec5}.

\subsection{Final size data}

Most disease outbreaks of concern, whether in human or animal populations, consist of many individuals getting infected, implying that by necessity the population size $n$ is also large. However, in veterinary science it also happens that controlled experiments are performed, where disease spread is studied in detail in several small isolated units (e.g.\ \cite{klinkenberg2002within}). We start by describing how to make inference in this situation, i.e.\ when observing disease spread in many small units. We do this for the somewhat simpler discrete time Reed-Frost model in which an infected individual infects other individuals independently with probability $p$. If we start with $k$ isolated \emph{pairs} of individuals, one being initially infected and the other initially susceptible, then $p$ is estimated by $\hat p=Z/k$, the observed fraction that were infected by the infected ``partner'' of the same isolated unit. This estimator is based on a binomial experiment and it is well-known that it is unbiased with a standard error of $s.e.(\hat p)=\sqrt{\hat p(1-\hat p)/k}$. A confidence bound on the estimator is constructed using the normal distribution and it is observed that the uncertainty in the estimator decreases with the number of pairs in the experiment as expected. Having estimated the transmission probability $p$ the natural next step is to estimate $R_0$. This is however non-trivial since moving the animal to its natural habitat in some herd will probably change the transmission probability $p$ (to each specific individual) to something smaller. If the transmission probability is the same when the individual is in its natural habitat, the basic reproduction number will equal $R_0=mp$ if there are $m$ individuals in the vicinity of any individual. This type of inference, for isolated units, can be extended to situations where there are more than two individuals out of which at least one is initially inoculated. However, the inference gets fairly involved even with very moderate unit sizes (e.g.\ size 4 units) due to the dependence between individuals getting infected. We refer the reader to e.g.\ \cite{becker1999statistical}, who also considers vaccinated and unvaccinated individuals with the aim to estimate vaccine efficacy, for further treatment of these aspects.

We now treat the situation when one large outbreak takes place in a large community (of uniformly mixing homogeneous individuals). As before, we let $n$ denote the population size and we assume data consists of the final size $Z=$ the ultimate number of infected individuals during the course of the outbreak. Using results from probabilistic analyses of a class of epidemic models (containing the general stochastic epidemic model) it is known that in case a major outbreak occurs in a large community, then the outbreak size $Z$ is approximately normally distributed with mean $n\tau$  and variance $n\sigma^2$ where $\tau$ and $\sigma^2$ are functions of the model parameters. These results, together with delta-method, can be used to obtain an explicit estimate $\hat R_0$ and standard error for the estimate (see  Section 5.4 in \cite{diekmann2012mathematical}):
\begin{equation*}
\hat R_0=\frac{-\log (1-Z/n)}{Z/n} \qquad s.e.(\hat R_0)=\frac{1}{\sqrt{n}}\sqrt{ \frac{1+c_v^2(1-Z/n)\hat R_0^2}{(Z/n)(1-Z/n)}} . \label{R_0fin_size}
\end{equation*}
The point estimate is based on the so-called final size equation  for the limiting fraction infected $\tau$: $1-\tau=e^{-R_0\tau}$. The expression for the standard error contains one unknown parameter $c_v$ which is the coefficient of variation of the duration of the infectious period $T_I$: $c_v^2=V(T_I)/(E(T_I))^2$. For the general stochastic epidemic the infectious period is exponential leading to that $c_v=1$ whereas $c_v=0$ for the Reed-Frost epidemic. Most infectious diseases have an infectious period with less variation than the exponential distribution, so replacing $c_v$ by 1 usually gives a conservative (i.e.\ large) standard error.

In case the outbreak takes place in a large community it may be that the total number infected $Z$ is not observed, but instead the number of infected $Z_m$ in a sample of size $m$ (say) may be the data at hand. Then there are two sources of error in the estimate: the uncertainty from the final outcome being random, and the uncertainty from observing only a sample of the community. The latter is of course bigger the smaller sample is taken. In this situation, the estimator of $R_0$ and its uncertainty are given by
\begin{align*}
\hat R_0 &=\frac{-\log (1-Z_m/m)}{Z_m/m} 
\\
s.e.(\hat R_0) &=\sqrt{ \frac{1+c_v^2(1-Z_m/m)\hat R_0^2}{n(Z_m/m)(1-Z_m/m)}    + \frac{(1-m/n)(1-(1-Z_m/m)\hat R_0)^2}{m(Z_m/m)(1-Z_m/m)}} . \label{R_0fin_size_sample}
\end{align*}
The above approximation uses the delta-method together with the fact that $V(Z_m)=E(V(Z_m|Z))+V(E(Z_m|Z))$. We see that the first term in the square root equals the standard error when observing the whole community and the second term vanishes if $m = n$ as expected. If on the other hand $m\ll n$ the second term under the square root dominates; then nearly all uncertainty comes from observing only a small sample.

Another fundamental parameter mentioned above is the critical vaccination coverage $v_c$: the necessary fraction to immunize in order to surely prevent a major outbreak. For our simple model we know that $v_c=1-1/R_0$. The estimator for this quantity is obtained by plugging in the estimator for $R_0$ given above, and a standard error is obtained using the delta-method again. The result is
\begin{equation*}
\hat v_c=1-\frac{1}{\hat R_0}=1-\frac{Z/n}{-\log (1-Z/n)} \qquad s.e.(\hat v_c)=\frac{1}{\sqrt{n}}\sqrt{ \frac{1+c_v^2(1-Z/n)\hat R_0^2}{\hat R_0^4(Z/n)(1-Z/n)}} . \label{v_cfin_size}
\end{equation*}
In case only a sample is observed we have the following estimator and standard error:
\begin{align*}
\hat v_c &=1-\frac{Z_m/m}{-\log (1-Z_m/m)} 
\\
s.e.(\hat v_c) &=\sqrt{ \frac{1+c_v^2(1-Z_m/m)\hat R_0^2}{n\hat R_0^4(Z_m/m)(1-Z_m/m)}    + \frac{(1-m/n)(1-(1-Z_m/m)\hat R_0)^2}{m\hat R_0^4(Z_m/m)(1-Z_m/m)}} . \label{v_cfin_size_sample}
\end{align*}
As when estimating $R_0$ the second term vanishes as $m\to n$ whereas it dominates if we have a small sample, i.e.\ $m\ll n$.

The above estimates were based on final size data from one outbreak assuming that all $n$ individuals were initially susceptible. In many situations there are also initially immune individuals when an outbreak occurs. Suppose as above that there are $n$ initially susceptible and $Z/n$ denotes the fraction infected among the initially susceptible, but that there were additionally $n_I$ initially immune individuals. Then the estimate $\hat R_0$ above is actually an estimate of the effective reproduction number $R_E=sR_0$, where $s=n/(n+n_I)$ denotes the fraction initially susceptible (just as if a fraction $1-s$ were vaccinated). The estimate of $R_0$ and $v_c$ (the fraction necessary to vaccinate assuming everyone is susceptible) are then given by the expressions above replacing $\hat R_0$ by $\hat R_0/s$. The corresponding standard errors are as before but dividing by $s$ for $\hat R_0$, and multiplying by $s$ for $\hat v_c$.

\subsection{Temporal data}

The estimates of the previous section were based on observing the final outcome of an outbreak, denoted $Z$. Quite often some temporal data, such as weekly reported cases, are also observed. This will improve inference for $R_0$ and $v_c$ as compared with final size data. However, for the simple scenario of the current section where there are no individual heterogeneities and where individuals mix uniformly, the gain from having temporal data is limited. In \cite{Andersson2000}, Exercise 10.3, the precision based on final size data is compared with the estimation precision from so-called complete data, meaning that the time of infection and time of recovery of all infected individuals are observed. Even with such very detailed data the gain in reduced standard error is only of the order 10-15\%  for some common parameter values. Since most temporal data is less detailed than complete data, but more detailed than final size data, the gain from such temporal data will be even smaller, say 5-10\%. A disadvantage with using temporal data in the analysis is that the estimators and their uncertainties are quite involved, using for example martingale methods, as compared to the rather simple estimators for final size data given above. Further, for some partial temporal data types it might even be hard to specify what is observed in terms of model quantities and estimators may therefore be lacking. For this reason we do not present estimators for temporal data and refer the interested reader to e.g.\ \cite{diekmann2012mathematical}, Section 5.4.

Having temporal data is hence not so important for precision in estimation of $R_0$ and the critical vaccination coverage $v_c$ when having a homogeneous community that mixes (approximately) uniformly. However, temporal data may be useful for many other reasons. Firstly, having temporal data enables estimation of the two model parameters $\lambda$ and $\gamma$ separately, and not only the ratio of the two $R_0=\lambda/\gamma$. Another important reason is that it may be used as model validation. It can for example happen that the close contact parameter ($\lambda$) changes over time, for example due to increasing precautions of uninfected individuals. Without temporal data such deviation from the model above cannot be detected. Similarly, if the community actually is heterogeneous in some way this will typically lead to a quicker decrease of incidence as compared to a homogeneous community. Another reason to collect temporal data is of course that it is not necessary to wait until the end of the outbreak before making inference. This is particularly important for new emerging outbreaks (see Section 4 below). Moreover, infectious diseases surveillance systems rely on the availability of temporal data for early outbreak detection and forecasting, as explained in Section 6.

\section{Heterogeneities} \label{sec3}

The model treated in the previous section assumed a community of homogeneous individuals that mix uniformly. Reality is of course not like that and various heterogeneities affect the spreading patterns of an infectious disease. The type of heterogeneities to consider will depend on both the type of community and the type of disease. Think for example of influenza and a sexually transmitted disease; for these two disease the relevant contact patterns clearly differ. Roughly speaking, heterogeneities can be divided into two different sorts, individual heterogeneities and mixing heterogeneities. These will be discussed below in separate subsections as they quite often require different methods of both modelling and statistical analysis.

\subsection{Individual heterogeneities}

Individual heterogeneities are individual factors which affect the risk of getting infected or of spreading the disease onwards. This can for example be age and/or gender, (partial) immunity or vaccination status. Such factors can often be used to categorize individuals into different types of individuals, and outbreak data will then be reported as final size (or temporal) data separately for the different cohorts. This type of data is often called a multitype epidemic outbreak. Final size data would then be to observe the number, or fraction, infected in the different cohorts. If there are $k$ groups we let the final fraction infected in each group be denoted by $\tilde \tau_1,\dots ,\tilde \tau_k$, and the known community fractions of the different groups are given by $\pi_1,\dots ,\pi_k$ (so $\pi_i$ is the community fraction of individuals being of type $i$). From this data we would like to estimate the model parameters $\{ \lambda_{ij},\gamma_i\}$; there is now a close contact  (=transmission) rate between all pairs of groups ($\lambda_{ij}/n$ is the rate at which an infectious $i$-individual infects a given susceptible type-$j$ individual, and a type-specific recovery rate ($\gamma_i$ is the recovery rate for $i$-individuals). In general we hence have $k^2+k$ model parameters whereas the data vector has dimension $k$. Clearly it will hence not be possible to estimate all parameters from final size data. In fact, it will not even be possible to estimate the basic reproduction number $R_0$ consistently, where $R_0$ is now the largest positive eigenvalue of the so-called next generation matrix $M$ with elements $m_{ij}=\lambda_{ij}\pi_j/\gamma_i$. An intuitive explanation to this result is easy to give for the situation where $\lambda_{ij}=\alpha_i\beta_j$, so the first factor is the infectivity of $i$-individuals and the second factor the susceptibility of $j$-individuals. By observing the final outcome of a multitype epidemic it is possible to infer which types are more susceptible to the disease, but it is less clear which types that are more infectious in case they get infected, and the latter affects $R_0$ equally much. The equations which to base parameter estimates on are the following (corresponding to the final size equations for the multitype epidemic model):
\begin{equation*}
1-\tilde\tau_j =e^{-\sum_i \lambda_{ij}\pi_i\tilde\tau_i/\gamma_i},\qquad j=1,\dots ,k.
\label{multitype_inf}
\end{equation*}
If the number of parameters are reduced down to $k$, or if some parameters are known, the $k$ equations above may be used to estimate the remaining parameters including $R_0$. Uncertainty estimates can also be obtained using probabilistic results of \cite{ball1993final}, but to derive them explicitly remains an open problem.

An important common particular type of multitype setting is where there are asymptomatic cases. For many infectious diseases certain infected individuals have no symptoms but may still spread the disease onwards. This situation is slightly different from the description above in that there are not two distinguishable types of individuals; it is only upon infection that individuals react differently and either become symptomatic or asymptomatic. The most challenging statistical feature is that the asymptomatic cases are rarely observed, i.e.\ it is only the symptomatic cases that are observed. In order to make good inference in this situation it is necessary to obtain information also about what fraction symptomatic cases there are, for example by testing for antibodies in a random sample in the community.

\subsection{Heterogeneous mixing}

Individuals are also heterogeneous in the way they mix with each other. In the simple model defined in the previous section it was assumed that individuals mix uniformly with each other, but reality is of course nearly always more complicated, which hence should be taken into account in modelling and statistical analysis. For human diseases there are mainly two types of mixing heterogeneities that have been accounted for: households and networks. The first and most important is the relevance of household structure for many diseases: for diseases like influenza the risk of transmitting to a specific household member is much higher than the risk of transmitting to a (randomly selected) individual in the community. This can be modelled by assuming a transmission rate $\lambda_H$ to each individual of the same household, and another ``global'' transmission rate $\lambda_G/n$ (of different order) to each individual outside the household. The effect of such additional transmission within household is that infected individuals will tend to cluster in certain households leaving other households unaffected (e.g.\ \cite{ball1997epidemics}), and the higher $\lambda_H$ is, the more will infected individuals be clustered. This can be used when inferring model parameters including reproduction numbers as illustrated by \cite{ball1997epidemics}, but also more recently in e.g.\ \cite{fraser2007estimating}.

For temporal data the two different transmission rates may be disentangled more directly by comparing the current fraction of infectives in a household whenever infection occurs (cf.\ \cite{fraser2007estimating}). For a model having constant infectious rates throughout the infectious period, the log-likelihood contribution relevant for estimating $\lambda_G$ and $\lambda_H$ equals
\[
 \sum_{i,j}\log[ S_i(t_{ij}-)(\lambda_HI_i(t_{ij}-) + \frac{\lambda_G}{n}I(t_{ij}-))] - \int_0^{t_{obs}}\lambda_H(\sum_iS_i(u)I_i(u)) + \frac{\lambda_G}{n}S(u)I(u) du,
\]
where $\{t_{ij}\}$ are the observed infection times in household $i$, and where $I_i(t)$ and $I_i(t-)$ denote the number of infectives in household $i$ at $t$ or just before $t$ respectively, and similar for $S_i(t)$ and $S_i(t-)$, and where (as before) $S(t)=\sum_i S_i(t)$ and $I(t)=\sum_i I_i(t)$ are the corresponding totals. This likelihood can be used (assuming the rare situation where infection times are actually observed) to infer the transmission parameters $\lambda_H$ and $\lambda_G$, i.e.\ it enables distinction between if most transmission is within or between households. If only final size data is available it is still possible to determine if most transmission takes place within or between households by fitting parameters to the final size likelihood  using recursive equations (cf.\ \cite{ball1997epidemics}). This method also enables estimation of a reproduction number $R_*$, which is both more complicated to interpret and a more complicated function of model parameters. A similar structure to households, having higher transmission within the groups than between, is that of schools and, for domestic animals, herds. These units are larger thus allowing some large population approximations such that each herd may have its own $R_0$. A complicated inference problem lies in estimating the contact rates between herds using transportation data (e.g.\ \cite{lindstrom2009estimation}).

A different type of mixing heterogeneity which has received a lot of attention in the modelling community over the last 10-15 years is where the community is treated as a social network and where transmission takes place only (or mainly) between neighbours of the network (e.g.\ \cite{newman2003structure}).  Both the structure of the network as well as the transmission dynamics taking place ``on'' the network are important for inferring the potential of an outbreak ($R_0$) and effects of various preventive measures. A big difference from the household setting just discussed is that usually the underlying network is rarely observed. At best, certain local properties of the network, such as the mean degree, the degree distribution, the clustering coefficient and/or the degree-degree correlation, may be known or estimated. From such local data more global structures determining the potential of disease outbreaks are usually not identifiable (cf.\ \cite{Britton15102013}). 

\subsection{Spatial models}\label{sec-sm}

Infectious disease epidemics in populations are inherently spatial  because infectious agents are spread by contact from an infectious host to a susceptible host that is located ``nearby''. Heterogeneity in space may play an important role in the persistence and dynamics of epidemics. For example, localised extinctions may be more common in smaller subpopulations whilst coupling between subpopulations may lead to reintroduction of infection into disease-free areas. Understanding the spatial heterogeneity has important implications in planning and implementing disease control measures such as vaccination. 

One way to account for spatial heterogeneity is to extend the general epidemic model by partitioning the population into spatial subunits of the hosts: nearby hosts are grouped together and interact more strongly than the ones that are further apart. These are the so-called meta-population models (or patch models) and they have been used also to investigate aspects of global disease spread in e.g.\ influenza \citep{cooper2006delaying}. A simple two-patch spatial model where hosts move between the two patches at some rate $m$ independent of a disease status would be as follows:

\begin{align*}
\frac{dS_1(t)}{dt}=&-\lambda S_1(t)I_1(t)/n+m(S_2(t)-S_1(t))
\\
\frac{dI_1(t)}{dt}=&\lambda S_1(t)I_1(t)/n-\gamma I_1(t)+m(I_2(t)-I_1(t))
\\
\frac{dS_2(t)}{dt}=&-\lambda S_2(t)I_2(t)/n+m(S_1(t)-S_2(t))
\\
\frac{dI_2(t)}{dt}=&\lambda S_2(t)I_2(t)/n-\gamma I_2(t)+m(I_1(t)-I_2(t))
\end{align*}

where $S_i$, and $I_i, i=1,2$ are the number of susceptible and infected individuals in the 2 patches respectively. The degree of mixing between groups can be specified, relaxing the assumption of uniform mixing of all
individuals. 

Time series data sets of infectious disease counts are now increasingly available with spatially explicit information. Some work has been done on time series susceptible-infected-recovered (TSIR) models \citep{finkenstadt2002stochastic} and its extensions as epidemic metapopulation model assuming gravity transmission between different communities \citep{xia2004measles, jandarov2014emulating}. According to a generalized gravity model, the amount of movement between the patches (communities) $k$ and $j$ is proportional to $n_k^{\tau_1}n_j^{\tau_2}/d_{jk}^\rho$ with $\rho,\tau_1,\tau_2>0$ and $d_{jk}$ is the distance between the patches where $n_k$ is the community $k$ size. The transient force of infection by infected individuals in location $j$ on susceptible in location $k$ is $m_{j\rightarrow k, t}\propto \frac{n_{k, t}^{\tau_1} I_{j, t}^{\tau_2}}{d_{jk}^\rho}$.

\section{Statistical analysis of emerging outbreaks} \label{sec4}

One of the most urgent problems in infectious disease epidemiology over the last decade has been to quickly learn about new diseases (or new outbreaks of old diseases). Examples include SARS \citep{lipsitch2003transmission,riley2003transmission}, foot and mouth disease \citep{ferguson2001foot}, H1N1-influenza, \citep{yang2009transmissibility,fraser2009pandemic} and, most recently, the Ebola outbreak in West Africa \citep{periods2014ebola}. A difference from the situation discussed above is that here, in order to identify efficient control measures, estimations are urgent \emph{during} the outbreak. It is not possible to wait until the end of the outbreak and use final size data to infer $R_0$ and related parameters. Instead inference has to be performed during the early growing stage of the outbreak. Beside having less data this also introduces the risk of producing biased estimates from the fact that individuals  that are infected during early stages of an outbreak are usually not representative for the community at large. As an example, the early predictions of the HIV outbreak in the 1980's predicted tens of millions of infected within a couple of years, predictions which turned out to be way too high. One partial explanation to this and similar situations is that in a heterogeneous community highly susceptible individuals will get infected early in the epidemic and if predictions are based on the whole community being equally susceptible as the initial group of infected the predictions will overestimate the final size.

As described in ealier sections, the basic reproduction number $R_0$ carries information about the potential of the epidemic and hence also how much preventive measures are needed to stop an outbreak. During an emerging outbreak, the data (such as weekly reports of new cases) carry information about the exponential growth rate $r$ of the epidemic (also known as the Malthusian parameter), so estimates of $r$ are easily obtained. However, there is no direct relation between $r$ and $R_0$; for example, a disease with twice as high transmission \emph{and} recovery rate has the same $R_0$ but larger growth rate $r$. It is the so-called \emph{generation time} that determines $r$, the generation time is defined as the time between infection of an individual to the (random) time of infection of one of the individuals he/she infects. The Malthusian parameter $r$ is defined as the solution to the  Lotka-Volterra equation
$$ 
\int_0^\infty e^{-rt}\mu (t)dt,
$$
where $\mu(t)$ determines the expected generation time and is defined as the average rate at which an infected individual infects new individuals $t$ time units after he/she was infected. The shape of $\mu(t)$ is very influential on the value $r$, and the duration and variation of the latent as well as infectious periods have a large impact on $r$, and thus on what can be inferred also about $R_0$ in an emerging epidemic outbreak. See \cite{wallinga2007generation} for more about the connection between $r$, the generation time and $R_0$.

In most emerging outbreaks the distribution $\mu(t)$ of the generation time is not known and inference methods are needed. However, very rarely infections times, end of latency periods and end of infectious periods are observed. Instead, some related events, such as onset of symptoms and end of symptoms are at best observed. The time between such successive observable events, e.g.\ the time between onset of symptoms of an infected and the time of onset of symptoms of one the individuals infected by him/her, is denoted the \emph{serial times}. As has been thoroughly investigated by Svensson (2007), generation times and serial times need not have the same distributions, the latter typically has more variation. As a consequence, even though inference about the serial times is possible from observable data it cannot be used directly to infer the generation time.

A final complicating matter when inferring $r$ and $R_0$ using data from an emerging outbreak is that the ``forward'' process generation time (or serial time) is often estimated from data on the corresponding ``backward'' process. By this is meant that infected individuals are contact traced backwards in time aiming at finding the infection time since of its infector (e.g.\ \cite{periods2014ebola}). This seemingly innocent difference has the effect that the observed ``backward'' intervals will typically be shorter than the corresponding ``forward'' (generation or serial) intervals because in a growing outbreak the transmitting event is often not so long back since there are many more potential infectors more recently (cf.\ \cite{scalia2010some}). If this bias is not accounted for, predictions based on the backward intervals will be biased in that the predicted number of weekly cases will be overestimated.

As just explained, there are several potential pitfalls when estimating $R_0$ and effects of preventive measures from an ongoing emerging outbreak, the reason being that the observed/estimable growth rate $r$ is not directly related to $R_0$ but only indirectly through the generation time, and the latter is sensitive to usually unknown latent and infectious period distributions. But suppose this complicating problem is somehow under control. Is then estimation of $R_0$ straightforward? The immediate answer is that heterogeneities in the community also play a role when inferring $R_0$ in an emerging outbreak. However, \cite{trapman14} show that for the most commonly studied heterogeneities such as multitype epidemics, network epidemics and household epidemics, their effect is very minor. More precisely, estimating $R_0$ assuming a homogeneous community when in fact it is a multitype epidemic gives exactly the correct estimate of $R_0$, estimating $R_0$ assuming a homogeneous community when in fact it comes from a (configuration) network epidemic makes the estimate of $R_0$ slightly biased from above (the conservative, ``better'' direction), and finally estimation of $R_0$ assuming homogeneity when the outbreak agrees with a household epidemic will make the estimate of $R_0$ close to the correct value and most often conservative. As a consequence, when the relevant heterogeneities make up a combination of the above heterogeneities the simpler estimate assuming homogeneity will slightly overestimate $R_0$, see \cite{trapman14} for more on this topic.

\section{Estimation methods (for partially observed epidemics)}\label{sec5}

As mentioned in Section \ref{sec2}, the main difficulty in estimating parameters for epidemic models is that the infection process is only partially observed and observed quantities may be aggregated in time (e.g.\ weekly or monthly).  Therefore, the likelihood may become very difficult to evaluate, especially when considering
temporal data, involving integration over all unobserved quantities, which is rarely analytically possible. Data imputation methods embedded into statistical inference techniques, such as the expectation-maximisation (EM) algorithm and Markov chain Monte Carlo (MCMC) have been used to estimate the unknown parameters in epidemic models.

The EM algorithm has been considered for epidemic inference problems by e.g.\ \cite{becker1997uses}. If we denote with ${Y}$ the observed data, with {Z} the augmented data (latent or missing) and with ${ \theta}$ the parameter (vector)  to estimate, the EM algorithm seeks to find the maximum likelihood estimate of the marginal likelihood by iteratively applying the following two steps: the E-step (expectation step) and the M-step (maximisation step). Once an initial parameter $\theta_0$ is chosen, the E-step and M-step are performed repeatedly until convergence occurs, that is until the difference between successive iterates is negligible. The E-step consists of computing the expected value of the complete data log-likelihood conditional on the observed data and the parameter estimate $\theta^{(t)}$ at iteration $t$, i.e.\  $Q( \theta| \theta^{(t)}) = \operatorname{E}_{ {Z}| {Y}, \theta^{(t)}}\left[ \log L ( \theta; {Y}, {Z}) \right] \,$ and the M-step requires maximising the expectation calculated in the E-step with respect to $\theta$ to obtain the next iterate. The latent data should be chosen such that the log-likelihood of the complete data is relatively straightforward. However, the evaluation of the expectation step can be rather complicated.

Data-augmented MCMC can be used to explore the
joint distribution of parameters and latent variables in a similar fashion. Especially in the Bayesian context, the approach is straightforward and it consists in specifying an ``observation level'' model $P({  Y}|{  Z, \theta})$, a  ``transmission level'' model $P({  Z}|{ \theta})$ and a prior $p({  \theta})$, as explained in details in e.g.\ \cite{auranen2000transmission}, resulting in 
$P({  Y},{  Z}, \theta)=P({  Y}|{  Z, \theta})P({  Z}|{ \theta})p({  \theta})$. One drawback with this approach is that it requires high memory for large-scale systems and in addition, designing efficient proposal distributions for the missing data may be challenging.  Therefore, applications of data augmentation in MCMC have been mainly concerned with the situation in which data arise from a single large outbreak
of a disease   \citep{gibson1998estimating, o1999bayesian} or data on small outbreaks across a large
number of households \citep{o2000analyses}.

For large epidemics in large populations, another option is to find analytically tractable approximations of the epidemic model. In epidemic time series data a natural choice is to approximate continuous-time models by discrete-time models \citep{lekone2006statistical}. An important constraint in those models is that one observation period must effectively capture one generation of cases. This may be achieved only if the generation time of the disease is equal to the length of observation periods, or is a multiple of it. In the latter case, the data must be further aggregated, which may lead to an additional loss of information. 
\cite{cauchemez2008likelihood} propose a statistical framework to estimate epidemic time-series data tackling the problem of temporal aggregation (and missing data),  by augmenting with the latent state at the beginning of each observation period and introducing a diffusion process that approximates the SIR dynamic and has an exact solution.

\cite{ionides2006inference} formulates the inference problem for epidemic models in terms of nonlinear dynamical systems (or state-space models) which consist of an unobserved Markov process $Z_t$, i.e.\ state process and an observation process $Y_t$. The model is completely specified by the conditional transition density $f(Z_t|Z_{t-1},   \theta)$, the conditional distribution of the observation process $f(Y_t|{ Y}_{t-1}, {  Z}_t,   \theta)=f(Y_t|Z_t,  \theta)$ and the initial density $f(Z_0|  \theta).$ The basic idea is to consider the parameter $  \theta$ as a time varying process $  \theta_t$, i.e.\ a random walk in $R^\theta$ so that $E(\theta_t|\theta_{t-1})=\theta_{t-1}$ and $Var(\theta_t|\theta_{t-1})=\sigma\Sigma$, because estimation is known to be easier in this setting. Then, the objective is to obtain estimate of $\theta$ by taking the limit as $\sigma\rightarrow 0$. The authors use
iterated filtering to produce maximum likelihood estimates within a Sequential Monte Carlo (SMC) framework.

A general technique that alleviates the problems generated by likelihood evaluation and that is growing in popularity in various scientific fields is the so-called  Approximate Bayesian Computation (ABC). ABC utilises the Bayesian paradigm in the following manner: if $M$ represents the model of interest, then
the observed data Y are simply one realisation from M, conditional on its (unknown) parameters $\theta$. For a given set of candidate parameters $\theta$, 
drawn
from the prior distribution, we can simulate a data set $Y'$ from M. If $\rho(s(Y'),s(Y))\le \epsilon$, where $\rho$ is a similarity metric, $s(\cdot)$ is a set of lower dimensional (approximately) sufficient summary statistics and $\epsilon$ is chosen small, then $\theta'$  is a draw from the posterior. ABC (or likelihood-free computation) can be used with rejection sampling  \citep{trevelyan2009inference}, MCMC  \citep{marjoram2003markov} or SMC routines \citep{toni2009approximate}. A general criticism of this method concerns the level of approximation generated by: the choice of metric $\rho$ and summary statistic $s$, the tolerance $\epsilon$ and the number of simulations to obtain estimates.

For stochastic models where simulation is time consuming, it
may not be possible to use likelihood-free inference. Learning about parameters in a complex deterministic or
stochastic epidemic model using real data can be thought of
as a ``computer model emulation/calibration'' problem \citep{farah2014bayesian}. Emulators are statistical approximations of a
complex computer model, which allows for simpler and faster computations. The estimation of epidemic dynamics can be carried out by combining a statistical emulator with reported epidemic data through a regression model allowing for model discrepancy and measurement error. Recent work in emulation and calibration
for complex computer models for fitting epidemic models include  \cite{jandarov2014emulating}, where a Gaussian process approximation is chosen to mimic the disease dynamics model using key biologically relevant summary statistics obtained from
simulations of the model at different parameter values.

\section{Statistical models for infectious diseases surveillance}\label{sec6}

Infectious disease data are often collected for disease surveillance purposes and information is typically available as incidence counts aggregated over regular time intervals (e.g.\ weekly). As a consequence, individual information is often lost. Also, the number of susceptible individuals in a population is rarely available. The typical goal in a surveillance setting is to monitor disease incidence to detect outbreaks prospectively. Due to the lack of detailed information mentioned above, this is rarely achieved by fitting epidemic stochastic models to data, i.e.\ by explicitly modelling the transmission process. 

Commonly the problem is formulated as statistical model for detecting anomalies (step increase) in univariate count data time series  $\{y_t, t=1,2,\ldots\}$. The first approach dates back to \cite{farrington1996statistical} who compared the observed count of reported cases in the current week with an expected number, which is calculated based on observations from the past, i.e.\ similar weeks from the previous years from a set of so-called reference values. An upper threshold is then derived so that an outbreak alarm is triggered once the current observation exceeds this threshold. At time $s$,  ${  {\bf y}_s}=\{y_t; t\le s\}$ the statistic $r(\cdot)$ is calculated on the basis of ${{\bf  y}_s}$ compared to a threshold value $g$. This results in the alarm time $T_a=min\{ s\ge 1: r({ {\bf  y}_s})>g \}$. Several variations/extensions of the Farrington's method exist, \citep{Salmon2014monit}, based on two steps: first, a Generalized Linear/Additive Model (Poisson or Negative Binomial) is fitted to the reference values, and then the expected number of counts $\mu_s$ is predicted and used (with its variance) to obtain an upper bound $g_s$. The alarm is raised if $y_s>g_s$.
Other model generalizations allow the detection of sustained shifts (not only step increases) through cumulative sum methods \citep{hohle2008count}. Applications are in both human and veterinary epidemiology, see e.g.\ \citep{kosmider2006stastistical}.

In some settings, infectious disease data are available at a finer geographical scale (cases are geo-referenced). In these situations the problem of spatio-temporal disease surveillance can be formulated in terms of point-process models \citep{diggle2005point}. The focus is predicting spatially and temporally localised excursions over a pre-specified threshold value for the spatially and temporally varying intensity of a point process $\lambda^\ast(x,t)$ in which each point represents an individual case. In \cite{diggle2005point}, the point process model is a non-stationary log-Gaussian Cox process in which the spatio-temporal intensity, has a multiplicative decomposition into two components, one describing purely spatial $\lambda^\ast_0(x)$ and the other purely temporal variation $\mu_0(t)$ in the normal disease incidence pattern, and an unobserved stochastic component representing spatially and temporally localised departures from the normal pattern $\psi(x,t)$. Hence, the spatio-temporal incidence is $\lambda^\ast(x,t)=\lambda_0^\ast(x)\mu_0(t)\psi(x,t)$ for $t$ in the prespecified observation period $[0, T], T > 0$, and observation region. Within this modelling framework, anomaly is defined as a spatially and temporally localised
neighbourhood within which $\psi(x,t)$ exceeds an agreed threshold, $g$, via the predictive
probabilities $p(x,s;g)=P(\psi(x,s)>c|$data until time $s)$. 

Statistical models as the above mentioned, can also be used for the study of spatio-temporal correlations and patterns explaining the statistical variability in incidence counts.  As a consequence of the disease transmission mechanism, the observations are inherently time and space dependent and appropriate statistical models have to account for such feature in the data. Geographic information can be available at different scales. For example, as in \cite{diggle2005point}, an entire region is continuous monitored and a (marked) point pattern model representation like the above, has a branching process interpretation allowing the calculation of the expected number of secondary infections generated by an infective within its range of interaction (proxy for $R_0$), see \cite{Meyer2014spatio}. A second possibility is that infections are obtained at a discrete set of units at fixed locations followed over time, as farms during livestock epidemics \citep{keeling2008modeling}. In this case, an SIR modelling approach can be pursued. A third case, probably the most common one, is to have individual data aggregated over some administrative regions and convenient period of time.

A general statistical framework for modelling data from the latter case can be found in \cite{paul2008multivariate} that extends the model previously proposed by \cite{held2005statistical}. The model is based on a Poisson branching process with immigration and can be seen as an approximation to a chain-binomial model without information on the number of disease susceptibles. Previous counts enter additively on the conditional mean counts that is decomposed in two parts: the {\it endemic} part and the {\it epidemic} part. The former explains a baseline rate of cases that is persistent with a stable temporal pattern, while the latter should account for occasional outbreaks. In particular, the number of cases observed at unit $i$ at time $t$, $i=1,\ldots,m$, $t=1,\ldots,T$ is denoted by $y_{it}$. The counts follow a Negative Binomial distribution $y_{it}|y_{it-1}\sim NegBin(\mu_{it},\phi)$ with conditional mean $\mu_{it}= \lambda' y_{it-1}+\exp(\eta_{it})$ and conditional variance $\mu_{it}(1+\phi \mu_{it})$ where $\phi>0$ is an overdispersion parameter and $\lambda'$ is an unknown autoregressive parameter. The {\it epidemic} component is represented by $ \lambda' y_{it-1}$ and the {\it endemic} part is $\exp(\eta_{it})$. The inclusion of previous cases allows for temporal
dependence beyond seasonal patterns within a unit. To explain the spread of a disease across units, the {\it epidemic} component can be formulated as $ \lambda' y_{it-1}+\gamma_i \sum_{j\neq i}w_{ji}y_{j,t-l}$ where $y_{j,t-l}$ denotes the number of cases observed in unit $j$ at time $t-l$ with lag $l \in {1,2,\ldots}$ and $w_{ji}$ are suitably chosen weights. To model seasonality, the {\it endemic} component can be specified as $\alpha_i+\sum_{s=1}^{S} \beta_s sin(\omega_st) +\delta_s cos(\omega_s(t))$ where $\omega_s$ are Fourier frequencies and the parameter $\alpha_i$ allows for different incidence levels in each of the $m$ units.

Statistical models for surveillance are commonly evaluated and selected in terms of predictive performance in one step ahead-prediction. Strictly proper scoring rules are generally used for this purpose \citep{gneiting2007strictly}, the most popular being the logarithmic score. A broad range of statistical methods for disease surveillance are implemented in the R package {\tt surveillance} \citep{hohle2013r}. 

\section{Concluding remarks}\label{sec7}

In this paper we have presented results for the general stochastic epidemic model and shown how to infer the most important epidemiological parameters, $R_0$ and $v_c$ under different data scenarios (final size data or temporal data). The general stochastic epidemic model assumes a finite population that mixes homogeneously and a constant infection rate $\lambda$ during the infectious period. In Sections \ref{sec3} and \ref{sec4} we have elaborated some model extensions, e.g.\ individual heterogeneity, heterogeneous mixing and spatial models discussing how estimation changes.

However, there are other features that affect the disease spread (and therefore other model extensions to account for them) that have not been treated in this work. For example, the probability of getting infected with a disease is usually not constant in time: some diseases are seasonal e.g.\ common cold viruses. Also an ``external'' change e.g.\ the implementation of a control measure, may affect either contact rates or infectiousness (or both). One way to account for that is to let the infection rate $\lambda$ change in time, e.g.\ as a periodic function \citep{cauchemez2008likelihood}.

Epidemic models can also be used to derive estimators for the efficacy of control measures such as vaccine, using data generated by field trials and observational studies. Understanding the relation between disease dynamics and interventions is essential particularly for vaccination programs. In fact, vaccines can have protective effects in reducing susceptibility, infectiousness or both and efficacy estimation has to be performed accordingly \citep{halloran2009design}.

In Section 6 we have discussed statistical models for infectious disease surveillance. Some other challenges in this area not treated in this work  include: under-reporting, differences in case definitions, zero inflation.

Over the last few years, an alternative approach for modelling infectious disease outbreaks has focused on phylodynamics, the integration of phylogenetic methods to analyze the genetic variation of the pathogen and epidemic models \citep{grenfell2004unifying}. This approach offers new insights into the dynamics of disease outbreak with the aim of inferring transmission routes and times of infection (see e.g.\ \cite{volz2009phylodynamics}).

\section*{Acknowledgments}

Both authors are grateful to the Swedish Research Council (grant 340-2013-5003) for financial support.

\bibliographystyle{apa}
\bibliography{bib1}

\end{document}